\documentclass[showpacs,preprintnumbers,amsmath,amssymb,twocolumn]{revtex4}

\usepackage{graphicx}
\usepackage{dcolumn}
\usepackage{bm}

\newcommand{\pr}{Phys. Rev. }

\newcommand{\Apj}{Astrophys. J. }

\begin{document}

\title{Heavy Element Production in Inhomogeneous Big Bang Nucleosynthesis}
\author{Shunji Matsuura
\footnote{E-mail: smatsuura@utap.phys.s.u-tokyo.ac.jp}}
\affiliation{Department of Physics, School of Science, University of Tokyo,
7-3-1 Hongo, Bunkyo, Tokyo 113-0033, Japan}

\author{Shin-ichirou Fujimoto }
\affiliation{Department of Electronic Control, 
Kumamoto National College of Technology, Kumamoto 861-1102, Japan}

\author{Sunao Nishimura}
\affiliation{Department of Physics, School of Sciences, 
Kyushu University, Fukuoka 810-8560, Japan}

\author{Masa-aki Hashimoto}
\affiliation{Department of Physics, School of Sciences, 
Kyushu University, Fukuoka 810-8560, Japan}

\author{Katsuhiko Sato}
\affiliation{Department of Physics, School of Science, University of Tokyo,
7-3-1 Hongo, Bunkyo, Tokyo 113-0033, Japan}
\affiliation{Research Center for the Early Universe, University of Tokyo, 
7-3-1 Hongo, Bunkyo, Tokyo 113-0033, Japan}

\begin{abstract}


 We present a new astrophysical site of the big bang nucleosynthesis (BBN) that
 are very peculiar compared with the standard BBN.
Some models of the baryogenesis suggest that very high baryon density
regions were formed in the early universe.
On the other hand, recent observations suggest that heavy elements
already exist in high red-shifts and the origin of these elements
become a big puzzle.
Motivated by these, we investigate BBN in very 
high baryon density 
regions. BBN proceeds in proton rich environment
, which is 
known to be the p-process like. 
However, by taking very heavy nuclei into account, we find that BBN proceeds through both the p-process
and the r-process simultaneously.
P-nuclei such as $^{92}$Mo, $^{94}$Mo, $^{96}$Ru, $^{98}$Ru whose origin is
not known well are also synthesized.

\end{abstract}


\maketitle

\section{INTRODUCTION}


What happened in the early universe has a great influence on the
history of the universe because they determined the initial conditions.
It is very important to check whether our standard model of cosmology is
correct or not as theories and observations develop.
Baryogenesis and BBN should be checked because they determine
 the history of the chemical evolution.

In the standard model of elementary particle physics,
the baryogenesis is possible only through the electro-weak spharelon
process. In the supersymmetric standard model, it is much easier 
to explain the baryon number asymmetry because there are many 
scalar fields which have baryon number.
One of the most striking property of supersymmetric theories
is that they have flat directions in potentials.
Some of them have baryon number and if fields
condensate in these directions, it is possible to
produce large baryon number.
This is the basic idea of the Affleck-Dine baryogenesis~\cite{affl-dine}.
Usually baryon number production is assumed to be taken place 
homogeneously all over the space.
This is natural because we know that the universe is homogeneous
and if baryogenesis is inhomogeneous in large scale 
it contradicts observations~\cite{cmb}.
Of course resolution ability of observations is limited and
small scale inhomogeneity is not excluded by observations.
Though it seems to be unnatural to consider 
such small scale inhomogeneity, 
recent observations force us to reconsider the possibility 
of inhomogeneous baryogenesis.

It has become clear that the evolution of matter started earlier than we have known before.
For example, Wilkinson Microwave Anisotropy Probe (WMAP) data suggests that
 reionization began when $z \sim$20 \cite{WMAP}.
According to Refs.~\cite{Barth:2003, Dietrich:2002}, 
 star formation activity 
started when $z \geq$  10. 
In addition, 
it is known that the quasar metallicity did not significantly  
change from the time of high redshift to 
the present time~\cite{Boksenberg:2003}.
Recently a galaxy at $z$=10.0 
was observed~\cite{Pello}.
Other evidences of heavy elements from the high redshifts are given
in
\cite{Pichon:2003, Cohen, Songaila, Prochaska, Freuding, Pettini}

 Motivated  by these observational evidences, we investigate the possibility
 that inhomogeneous baryogenesis produced very high baryon density in
 small fraction of the universe and in these regions
 some fraction of heavy elements were already synthesized during
 BBN.

 Heavy elements production during BBN itself
 is not a new idea.
 Previous researches on the inhomogeneous 
 big bang nucleosynthesis are given in~\cite{IBBN1, IBBN2}.
Heavy elements production is also mentioned in~\cite{jedam}.
These works, however, do not include very heavy elements~\cite{IBBN1}
or they create neutron rich regions and calculate the nucleosynthesis in
those regions~\cite{IBBN2}. These are not suitable for our present purpose.

 
 For the production of heavy elements during BBN,
 high baryon density is necessary.
 However if we simply increase the baryon-photon ratio all over the universe
 homogeneously, it would apparently contradict the observed light element
 abundances~\cite{light1} and CMBR~\cite{cmb}.
 Instead, we assume that the baryon density of the universe is inhomogeneous
 before and during BBN.
 In most part of the universe $\eta$ is
 small ($\eta \sim 6 \times 10^{-10}$) as observed
 while small fraction of the universe is occupied with very high baryon density,
 $\eta \sim \mathcal{O}(1)$. 
Because our aim is to see how BBN goes 
 in the high baryon density regions and not make the precise adjustment 
 between BBN and CMBR, we neglect the baryon diffusion. 
In this case, the baryon density in high density regions can be treated almost
free parameter without contradicting observations. 
(It is a complicated problem whether we can treat $\eta$ as a free parameter 
in realistic models.
See, for example,~\cite{neutrino}.)


In section~\ref{theo} we explain the theoretical aspects of our
model~\cite{Dolgov:1993si}.
In section~\ref{Numerical}, we explain our network and what kind of
effects we take into account.
Section~\ref{Results} is the main results of our numerical study.
In this section we explain BBN is the p-process like and simultaneously the
r-process like. And also BBN can produce very heavy elements including
proton rich nuclei such as $\rm ^{92}Mo, ^{94}Mo, ^{96}Ru, and ^{98}Ru$.

\section{THEORETICAL BACKGROUND}
\label{theo}

 
 Theoretical background of this model is inhomogeneous
 baryogenesis~\cite{Dolgov:1993si}. We are going to explain basic aspects
 of this model. For more detail, see~\cite{Dolgov:1993si, matsu}.

The basic idea of the model \cite{Dolgov:1993si} is a modified version of the
Affleck-Dine baryogenesis~\cite{affl-dine}.
Assume that the interaction Lagrangian has the general renormalizable form
\begin{equation}
 	\begin{split}
    	 \mathcal{L}_{int} &=\lambda |\phi|^2 \Phi ^2 + g |\phi|^2 \Phi \\
	    	&= \lambda (\Phi - \Phi _1)^2 |\phi|^2 - \lambda \Phi ^2_1|\phi|^2 ,
  	\end{split}
\end{equation}		
where $\phi$ is Affleck-Dine (AD) field, $\Phi$ is the inflaton field, 
$g$ and $\lambda$ are the coupling constants and $\Phi _1 = -g/2\lambda$.

In a simplest case, the effective mass of the AD field $\phi$ can be written by
\begin{equation}
(m^{\phi}_{eff})^2=m^2_{0}+\lambda (\Phi - \Phi _1)^2 ,
\end{equation}
where $m^2_{0}$ is the vacuum mass of $\phi$.

The vacuum expectation value of $\Phi$ is assumed to evolve from 
very large value, i.e. , $\Phi \geq \Phi _1$, decreases to zero.
As $\Phi$ goes down to $\sim \Phi _1$,  the effective mass square of $\phi$
becomes negative and the phase transition takes place. When 
$\Phi$ is far from $\Phi _1$ the mass square is positive.
If the duration of $\Phi \sim \Phi _1$ is short,
the transition would take place only in a small fraction of space.
Consequently in the dominant part of the universe baryon asymmetry is small
as observed $\eta = \mathcal{O}(10^{-9})$, while in a small part of the universe
 the baryon asymmetry can be very large, even close to unity.
In this simple model, the signature of barionic charge is not fixed.
Baryonic chaege can become both positive and negative \cite{B-charge}.
However, the high density regions are very small compared to 
cosmological scale, high density anti-matter regions would
disappear by pair-annihilation while late time inflation
can prevent the annihilation \cite{domain}.

 Because we are not very interested in the detail of the shape
 of the bubbles and the effect of diffusion in this paper,
 we assume that the
 bubble sizes are large enough to neglect the diffusion effects.
 Also, bubbles are not large so as to contradict 
 the observations~\cite{cmb}.

 In this case, the BBN calculation
 can be treated as that of homogeneous big bang nucleosynthesis.
 In the following section, we present the results of the calculations and 
 their physical interpretations.


At first sight, in the high baryon density regions the reaction seems to proceed along the proton rich side 
because BBN occurs in proton rich environment~\cite{matsu}.
However, surprisingly it is not correct. BBN proceeds along the
{\bf proton rich} and the {\bf neutron rich} side.

\section{NUMERICAL CALCULATIONS}
\label{Numerical}

The basic method of our calculation is the same as that of the homogeneous big bang nucleosynthesis.

We solve the Friedmann equation 
\begin{equation}
\left(\frac{\dot{a}}{a}\right)^2 =\frac{8\pi G \rho}{3}
\end{equation}
where $\rho = \rho _{\gamma} + (\rho _{e^-} + \rho _{e^+} ) + \rho _{\nu} + \rho _{b}$, and
$a$ is the scale factor.
The energy conservation law is
\begin{equation}
\frac{d}{dt}(\rho a^3) + \frac{p}{c^2}\frac{d}{dt}(a^3)=0
\end{equation}
for the time evolution of the temperature and the baryon density.

Abundance change in the region is evaluated with a nuclear reaction network, 
which includes 4463 nuclei from neutron, proton to Americium ({\it Z} = 95, {\it A} = 292).
Nuclear data, such as reaction rates, nuclear masses, and partition functions, 
are same as in~\cite{fujimoto}.
It should be emphasized that both proton-rich and neutron-rich nuclei are 
produced in a high density region (Fig.~\ref{600}).
Therefore it is required using a large network to calculate abundances in the
high $\eta$ region.

\section{RESULTS}
\label{Results}

\begin{figure}[htbp]
\begin{center}
\includegraphics[height=8cm,width=9cm,clip]{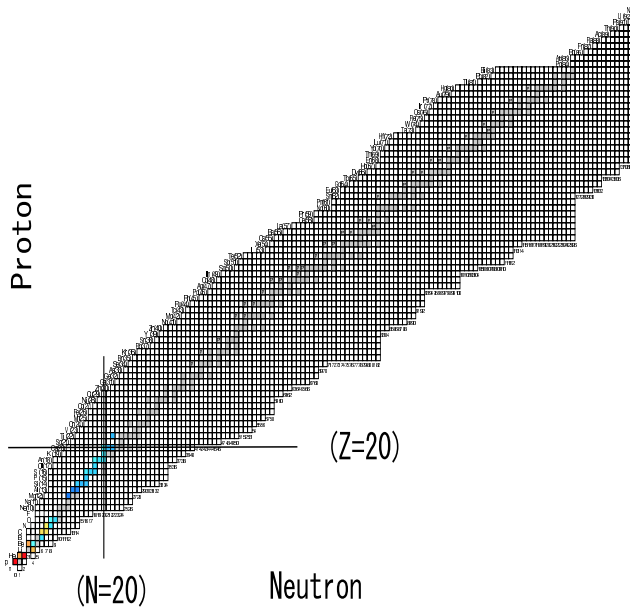}
\caption{Abundance distribution at ${\it T} = 1 \times 10^{7}$K ($\eta = 10^{-6}$) is plotted
on the nuclear chart. The gray color regions represent stable nuclei.
The red color represents more synthesized nuclei and blue color
represents less synthesized nuclei.
Those nuclei that are not synthesized in the standard BBN scenario, such as
Ca are synthesized. }
\label{600}
\end{center}
\end{figure}

We have calculated BBN for various values of $\eta$, from 
$10^{-10}$ to $10^{-2}$.
It is known that in the standard, low baryon density BBN, 
nuclei heavier than Boron are hardly synthesized.
Fig.~\ref{600}  
represents the synthesized nuclei for $\eta=1\times 10^{-6}$ at the epoch $T=1\times 10^{7}$K.
We can see that heavier nuclei such as Ca ($10^{-14}$ in mass fraction) are synthesized.

Naturally as $\eta$ becomes large, heavier nuclei are synthesized.
However we find there is drastic change in the nucleosynthesis around 
$\eta \sim 3 \times 10^{-4}$.
To see this transition, we pick up two values of $\eta =10^{-4}$ and $10^{-3}$,
and investigate
what is going on during the nucleosynthesis.

\begin{figure}[htbp]
\begin{center}
\includegraphics[height=8cm,width=9cm,clip]{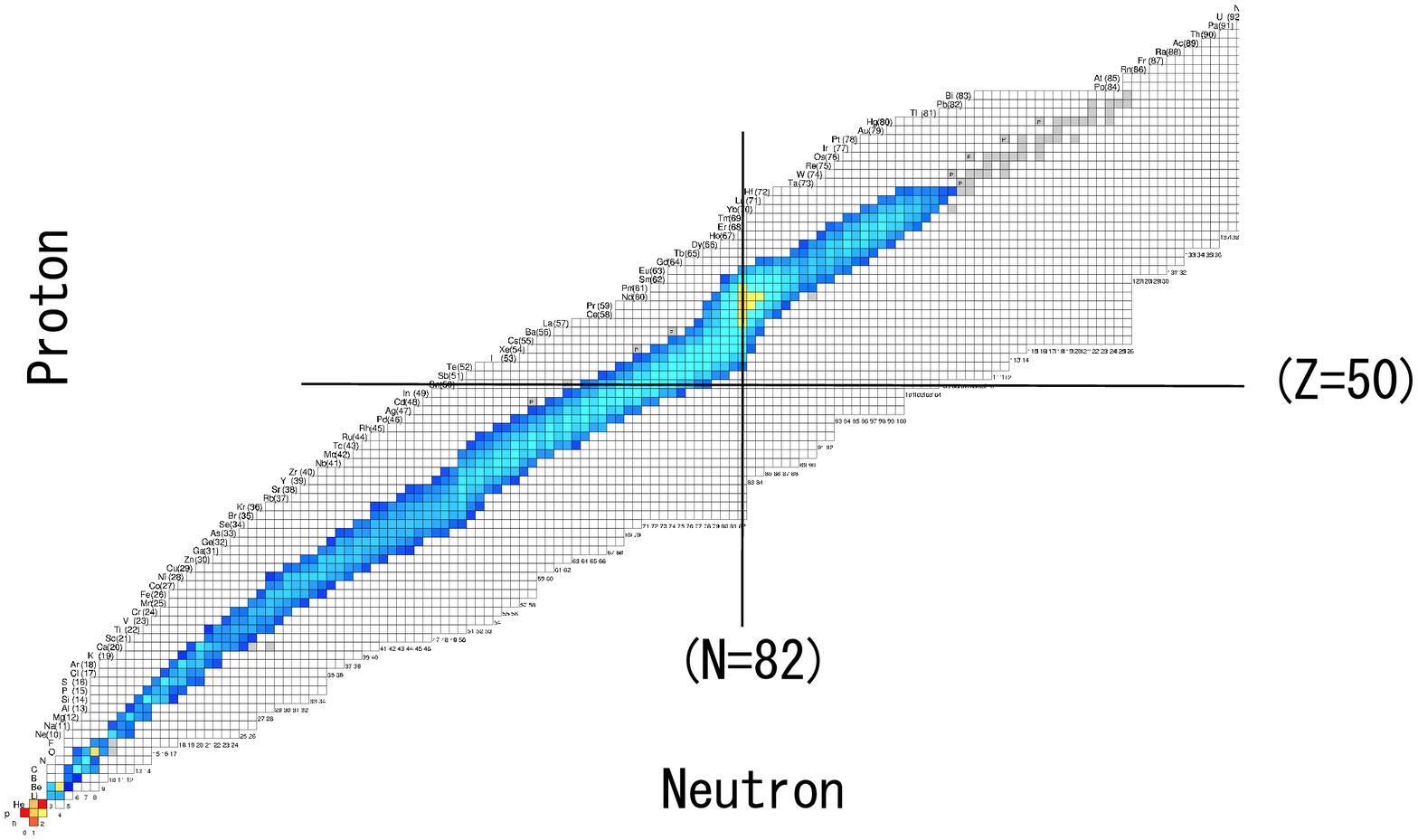}
\caption{Abundance distribution at ${\it T} = 3 \times 10^{9}$K ($\eta = 10^{-4}$) are plotted
on the nuclear chart. The red color represents more synthesized nuclei. 
Produced nuclei distribute along the stable nuclei.}
\label{mo1}
\end{center}
\end{figure}

Fig.~\ref{mo1} represents how many nuclei are synthesized 
at the temperature ${\it T} = 3 \times 10^{9}$K, and
$\eta = 10^{-4}$.

\begin{figure}[htbp]
\begin{center}
\includegraphics[height=8cm,width=9cm,clip]{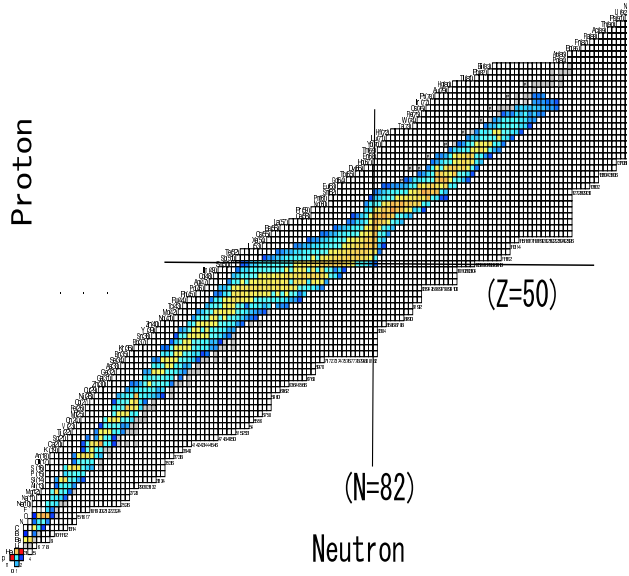}
\caption{Abundance distribution at ${\it T} = 1 \times 10^{9}$K ($\eta = 10^{-4}$). 
For nuclei with ${\it A}$ $\leq$ 100 the synthesis is the p-process 
while for heavier nuclei the synthesis is the r-process.
}
\label{mo2}
\end{center}
\end{figure}

\begin{figure}[htbp]
\begin{center}
\includegraphics[height=8cm,width=9cm,clip]{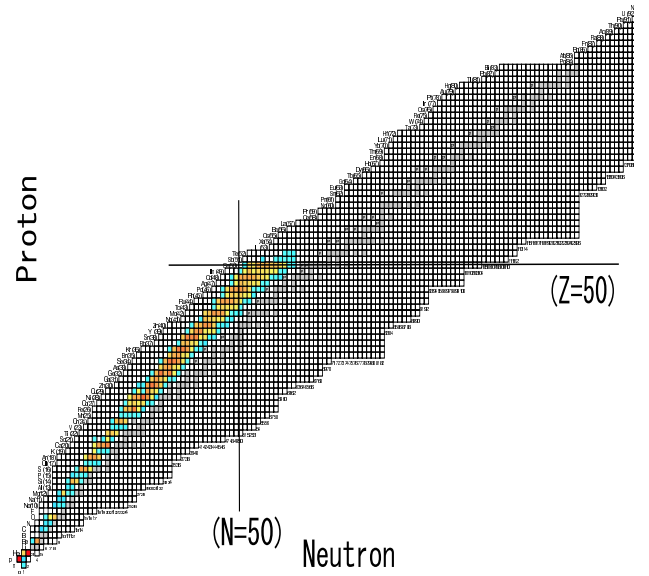}
\caption{Abundance distribution at ${\it T} = 1.8 \times 10^{9}$K ($\eta = 10^{-3}$).
Nucleosynthesis occurs through the p-process and very heavy nuclei are not synthesized.
}
\label{mo4}
\end{center}
\end{figure}

Red (blue) color represents more (less) synthesized nuclei.
(Stable nuclei also plotted in Fig.~\ref{mo1},~\ref{mo2}, and~\ref{mo4}, with gray color.)
We can see that the reaction goes along stable line.
It is well known that nuclei whose neutron and  proton numbers are
special values (for example 20, 28, 50, 82 etc.), the magic numbers, are especially stable.
At these points, the reactions are stagnated and the reaction paths are bent.
Especially, the stagnation at ${\it N}$ = 82 is one of the biggest factors that
 prevent the reaction to proceed further beyond the
mass number 190.
As the temperature goes down, the locus of the reaction 
begins to bent to different directions (Fig.~\ref{mo2}).

For lighter nuclei (mass number $A  \leq 100$),
proton captures are very active and the locus moves to 
proton rich direction.
For nuclei whose mass numbers are between 100 and 120,
the locus is across the stable nuclei from proton rich side to
neutron rich side.
For heavier nuclei for ${\it A} \ge 120 $,
neutron capture is more efficient.
This suggests that both the r-process and p-process occur simultaneously in BBN.

\begin{figure}[htbp]
\begin{center}
\includegraphics[height=8cm,width=9cm,clip]{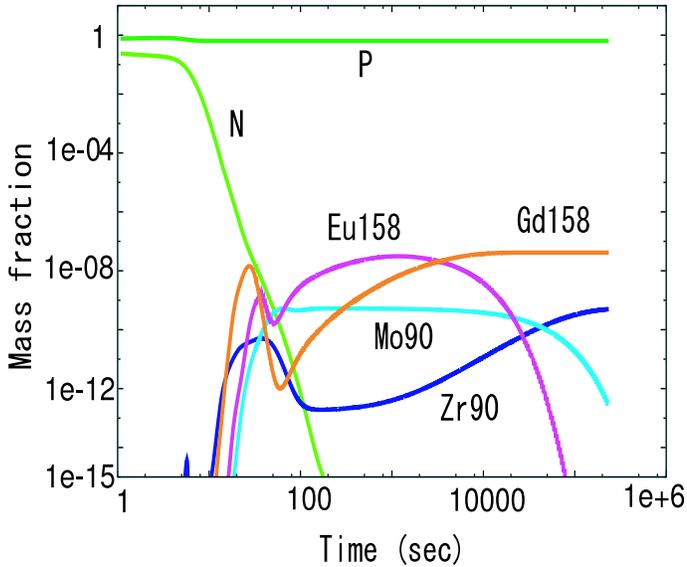}
\caption{Time evolution of mass fraction ($\eta=1\times 10^{-4}$). Neutrons are still left when
heavy elements are synthesized. For nuclei with mass number 90,
stable ones are synthesized first followed by proton rich nuclei. On the other
hand, for nuclei with mass number 158 stable nuclei are synthesized first followed by
 r-rich nuclei.}
\label{mo3}
\end{center}
\end{figure}

 Physical interpretation of this situation is as follows.
 The environment in BBN is proton rich, i.e.,    
the electron fraction ${\it Y}$e ranges from 0.8 to 0.9~\cite{matsu}.
 Naive expectation of BBN is the
 p-process.
 For relatively light heavy nuclei, proton capture is active.
 However proton capture processes become exponentially difficult as the
 proton number increases because of their coulomb barrier.
 On the other hand, neutrons are still not consumed out during 
 heavy nuclei are synthesized as shown in Fig.~\ref{mo3}.
 Very heavy nuclei captures neutrons and the locus of the
 reaction changes toward the neutron rich side.
 Transition point from proton rich side to neutron side depends on
 the baryon-photon ratio $\eta$.
 The transition occurs at larger mass number for larger $\eta$.
 The reactions depend on the abundances of the seed nuclei.
 The higher baryon density follows many seeds which lead
 to proton captures on heavier nuclei.
 Fig.~\ref{mo3} shows the time evolution of mass fraction for nuclei whose mass
 number is 90 and 158.
 Let us see the time evolution of the mass number 90.
 $^{90}$Zr is a stable nucleus and $^{90}$Mo is a proton rich one.
 First, $^{90}$Zr is synthesized and later $^{90}$Mo is synthesized while the amount of $^{90}$Zr decreases.
 This represents that the reaction first proceeds along the stable line and later
 move to the proton rich side.
 At the late stage, $^{90}$Zr increases while $^{90}$Mo decreases because unstable 
 proton rich nuclei  decay to stable nuclei such as
 $^{90}$Mo $\rightarrow$ $^{90}$Nb $\rightarrow$ $^{90}$Zr.
 The lifetimes of $^{90}$Nb and $^{90}$Mo are 
 $14.6$ h = $5.33\times 10^{4}$ sec and $5.67$ h = $2.04\times 10^{4}$ sec, 
 respectively.

 For heavier nuclei of the mass number 158, the situation is different.
 $^{158}$Gd is a stable nucleus and $^{158}$Eu is a neutron rich nucleus, instead of 
 proton rich one.
 First the stable nucleus $^{158}$Gd is synthesized and later neutron rich $^{158}$Eu is synthesized.
 This shows that in heavier nuclei region, the stable nuclei are produced first as $^{90}Zr$, 
  but later the neutron rich nuclei are produced instead of proton rich nuclei.
  The decrease in $^{158}$Eu in the late stage is the same as $^{90}$Mo, $\beta$ decay to 
  stable nuclei.
  The abundances of neutron rich nuclei of the mass number 90 and those
  of proton rich nuclei of the mass number 158 are very small and not
  drawn in this figure.
  We can also see that neutrons are still left when heavy nuclei are synthesized.

\begin{figure}[htbp]
\begin{center}
\includegraphics[height=6.5cm,width=9cm,clip]{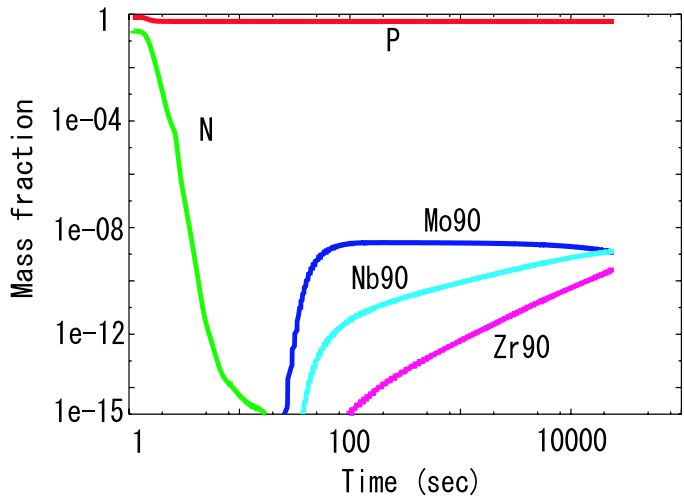}
\caption{Time evolution of mass fraction ($\eta = 1\times 10^{-3}$). Neutrons have already been consumed when
the abundances of heavy elements are increasing.}
\label{mo5}
\end{center}
\end{figure}

  Now let us see the case $\eta$ =$10^{-3}$.
Fig.~\ref{mo4} shows the locus of the reaction at the temperature  ${\it T} =1.8 \times 10^{9}$ K.
It is apparently different from the results of $\eta$= $10^{-4}$.
For, in this case, the reactions first proceeds along the stable line.
However, the reactions directly proceeds to the proton rich region.
Another important difference is that very heavy nuclei of ${\it A} \ge 80$ are not synthesized.
The physical interpretation is as follows.
In a high baryon density region, the seeds for the reactions to proceed are abundant.
The nuclear reaction proceeds promptly and all neutrons are consumed by
light nuclei as shown in Fig.~\ref{mo5}.
This prevents the nucleosynthesis from proceeding to the large 
mass number region.
In Fig.~\ref{mo5} we only draw the abundance
having ${\it A} = 90$.
Heavier nuclei are not synthesized enough.
When heavy nuclei are synthesized, neutrons are almost consumed out.

\begin{figure}[htbp]
\begin{center}
\includegraphics[height=7cm,width=9cm,clip]{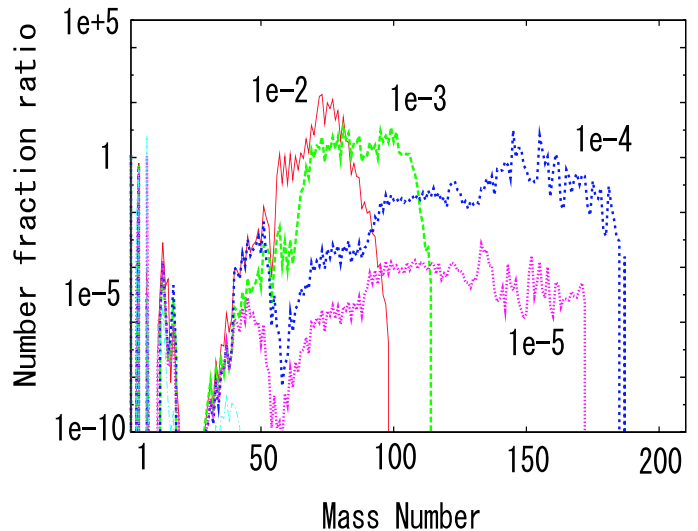}
\caption{The number fraction relative to the solar abundances.
As $\eta$ becomes larger, the maximum ${\it A}$ becomes larger for 
$\eta \leq 1 \times 10^{-4}$, while for  $\eta \geq 1 \times 10^{-4}$,
the maximum ${\it A}$ becomes smaller.}
\label{mo6}
\end{center}
\end{figure}

In Fig.~\ref{mo6}, we show the relation between the mass number and the number fraction relative to the solar abundances.
As the baryon density becomes larger, the heavier nuclei are synthesized for $\eta$ less than
$1 \times 10^{-4}$.
However, when $\eta \geq 1\times 10^{-4}$, the maximum mass number decreases as 
$\eta$ becomes larger.

The number fraction ratios of p-nuclei to the solar abundances is listed
in Table.~\ref{oo}.

\begin{table}[htbp]

\begin{tabular}{c|c|c|c}
\hline
$\eta = $ &$10^{-4}$  & $10^{-3}$ & $10^{-2}$ \\
\hline
$^{92}$Mo & $1.0\times 10^{-2}$ &  $1.1\times 10$  & $1.1\times 10^{-2}$ \\
$^{94}$Mo & $4.2\times 10^{-2}$   &  $0.9 \times 10 $   & $9.5\times 10^{-4}$ \\
$^{96}$Ru &   $9.6\times 10^{-2}$ & $3.1 \times 10$  & $ 8.1\times 10^{-5}$  \\
$^{98}$Ru &   $3.1\times 10^{-1}$ & $6.6\times 10$ & $3.5\times 10^{-6}$  \\
\hline

\end{tabular}
\caption{The number fraction of p-nuclei relative to the solar abundances.}
\label{oo}
\end{table}

For $\eta =10^{-3}$, $^{92}$Mo,  $^{94}$Mo, $^{96}$Ru and $^{98}$Ru drastically increase, 
due to the change of the loci of the reaction flows.
This suggests that highly inhomogeneous BBN would have a large influence
on the abundances of solar p-nuclei.


The observed abundances seem not to be explained
by BBN with only a single $\eta$.  
However this does not exclude the possibility of our assumption.
It is unnecessary for BBN abundances to match exactly to
the solar abundances because produced nuclei in high $\eta$ regions would have mixed with
nuclei synthesized in low density regions and also
there should be nuclei synthesized in star activities.


Basic feature of BBN at each $\eta$ are classified as follows.
For $\eta \leq 10^{-6}$, synthesized nuclei are limitted to ${\it A} \le 40$.
Nuclei whose ${\it A}$ are around 20 are less synthesized
even at large value of $\eta$.
For $\eta$ from $10^{-5}$ to $10^{-4}$, the abundances of nuclei whose 
mass number of $30 \le {\it A} \le 56$ grow
rapidly with ${\it A}$.
Abundances of nuclei ${\it A} \ge 56$ suddenly decrease but
again slowly increase. At around ${\it A} =$ 140, they turn to decrease.

After the rapid decrease in the abundances of ${\it A} \ge 56$ for 
$\eta = 10^{-3}$ and $10^{-2}$, the abundance profiles are rather different.

For $\eta=10^{-3}$, the abundances do not drastically change
from ${\it A} =$ 64 to around 86. They rapidly decrease for ${\it A}$  
above 100 and maximum ${\it A}$  synthesized is 114.
For $\eta=10^{-2}$, right side wing of Fe peak is similar to the solar
abundances. There is a peak around ${\it A} =$  72 and the abundance production 
decreases rapidly above beyond the peak until ${\it A} \sim 98$.
The maximum ${\it A}$ synthesized is 98.

To compare our results with observations such as metal poor star
abundances, we need to
take into account dynamical mixing after the epoch of BBN.
This depends on a model significantly and will be a future work.

We should examine the idea presented in this paper with
more realistic model, and determine whether
heavy elements were really synthesized in BBN or not.
The former is to take into account the
diffusion effects before and during BBN and also lepton asymmetry.
The latter  is to calculate the nucleosynthesis
in supermassive stars.
This is because supermassive stars are generally thought to 
have synthesized first heavy elements in the universe.
We need to know whether heavy elements observed in high redshift
were synthesized in BBN or supermassive stars.
Nucleosynthesis in supermassive stars and BBN in high
baryon density region is similar.
It would be a problem how to distinguish
these two nucleosynthesis from observations.
\par

\section{Conclusion}
We have investigated BBN in high baryon density region.
In these regions, not only light elements which are
synthesized in standard BBN but also very heavy elements
are produced.
We found BBN is both the p-process like and the r-process like.
The transition from the p-process to the r-process is due to
the Coulomb barriers of proton-rich nuclei 
and the amounts
of neutrons when heavy elements begin to be synthesized.
The loci of the reaction flows change drastically above
$\eta = 10^{-3}$. 
Above $\eta =10^{-3}$, a lot of seed nuclei cause active proton capture
and the reaction flows end before very heavy elements are synthesized.

Our calculations demonstrate that very heavy elements can be synthesized
in BBN, including proton-rich nuclei.
These nuclei will be related to the origin of the solar abundances, heavy elements
observed in high redshifts and early star formations via cooling effects.

For more realistic models in BBN, we need to include diffusion effects.
Comparison with the nucleosynthesis in supermassive stars
is also important.
We leave these issues for future study.

\section{Acknowledgements}
S.M. thanks Kazuhiro Yahata, Yuuiti Sendouda, Shigehiro Nagataki,
Naoki Yoshida, Mamoru Shimizu, Kohji Yoshikawa, Atsunori Yonehara, 
Shinya Wanajo, Tomoya Takiwaki,
Naoyuki Itagaki, Koji Higashiyama, Satoshi Honda for
useful discussions.
This research was supported in part by Grants-in-Aid for Scientific
Research provided by the Ministry of Education, Science and 
Culture of Japan through Research Grant No.S 14102004, No.14079202. 
S.M.'s work was supported in part by JSPS(Japan Society for the
Promotion of Science).

%
%


%

\end{document}